\magnification1200
\baselineskip=12truept
\vsize=22 truecm
\hsize=16.5 truecm
\parskip=0.2cm
\parindent=0.2truecm
\def\c{\centerline}
\def\noi{\noindent}
\def\bs{\bigskip}
\def\ms{\medskip}

\tolerance=10000
\def\today{\ifcase\month\or
January\or February\or 
March\or April\or May\or June\or
July\or August\or 
September\or October\or November\or 
December\fi\space\number\day, \number\year}

\newcount\fcount\fcount=0
\def\ref#1{\global\advance\fcount by 1 
\global\xdef#1{\relax\the\fcount}}

\def\pp{\parshape 2 0truecm 15truecm .5truecm 
14.5truecm}
\def\ref #1;#2;#3;#4{\par\pp #1, {\it #2}, {\bf #3}, 
#4}
\def\book #1;#2;#3{\par\pp #1, {\it #2}, #3}
\def\rep #1;#2;#3{\par\pp #1, #2, #3}

\def\simlt{\lower.5ex\hbox{$\; \buildrel < \over \sim 
\;$}}
\def\simgt{\lower.5ex\hbox{$\; \buildrel > \over \sim 
\;$}}

%
%
%


\newif\ifpsfiles\psfilestrue
\newcount\eqnumber
\eqnumber=0
\def\neweq{\global\advance\eqnumber by 1 (\hbox{\the\eqnumber})}
\def\eqnam#1{\xdef#1{(\the\eqnumber)} \relax
	     \immediate\write1{\def\string#1{#1}} }

\newcount\fcount \fcount=0
\def\ref#1{\global\advance\fcount by 1 
  \global\xdef#1{\relax\the\fcount}}

%

%
\def\pp{\parshape 2 0truecm 15truecm 2truecm 13truecm}
\def\apjref#1;#2;#3;#4 {\pp #1, {\it #2}, {\bf #3}, #4.}
\def\book#1;#2;#3 {\pp #1, {\it #2}, #3}
\def\rep#1;#2;#3 {\pp #1, #2, #3}
\def\unpub#1;#2 {\pp #1, {\it #2}}

\vskip -3cm
\today
\hfill CfPA-96-th-17

\bs\bs
\c{ \bf FEEDBACK, DISK SELF-REGULATION AND GALAXY FORMATION}
\bs\bs

  \c {\bf Joseph Silk}
  \c {Departments of Astronomy and Physics, and Center for Particle
  Astrophysics,}
  \c{University of California, Berkeley, CA 94720}
  \bs
  \bs
  \bs

  \c{\bf ABSTRACT}
\bs
Self-regulation of star formation in disks is controlled by two dimensionless parameters: the Toomre parameter for gravitational instability and the porosity of the interstellar medium to supernova remnant-heated gas. An interplay between these leads to expressions for the gas fraction and star formation efficiency in disks, and to a possible
explanation of the Tully-Fisher relation.
I further develop feedback arguments that arise from the impact of massive star
formation and death on protogalaxies in order to account for the characteristic luminosity of a galaxy and for early  winds from forming spheroids.
\bs\bs
\noi\c{ \bf 1. Introduction}
\medskip

 Star formation in disks appears to be self-regulated.  Cloud aggregation and star formation is controlled by the gravitational instability of a cold disk.
The Toomre parameter,
which controls  the growth rate of gravitational instabilities,
is near unity as a function of galactocentric radius,
within a critical gas surface density. 
Gravitational instability drives cloud aggregation and star formation, yet 
the star formation efficiency in disks is low, allowing the disk gas supply to be long-lived. The porosity of the interstellar medium to supernova remnant-heated gas is significant, and of order unity.
Porosity must evidently counter star formation. I argue that there is an anticorrelation  betweeen these two dimensionless parameters,  the gravitational instability parameter
and the porosity, that results in the  self-regulation of star formation.

A semi-phenomenological theory exists for star formation in disk galaxies.
 I show in Section 2 that the gas scale-height  is controlled by
the porosity of the interstellar medium in such a way that the gas velocity dispersion is constant.  Incorporation of the Toomre parameter for gravitational instability allows one to account for 
the inferred self-regulation of the star formation rate.
 An interplay betweeen   porosity  and gravitational instability
leads to a tentative
explanation of the Tully-Fisher relation, in which there is no explicit 
dependence on dark halos (Section 3).

Star formation in spheroids has a  far less secure foundation in theory and in phenomenology than star formation in disks. Indeed, one could safely say that there is essentially no theory and little in the way of phenomenology.
Recourse must be had to relatively crude scaling arguments that center on attempts to account for 
the origin of the galaxy luminosity function. This 
constitutes one of the outstanding problems in galaxy 
formation theory.  For example,
hierarchical merging of dark matter 
halos yields too steep a slope for the resulting 
luminosity function if mass traces light.  However, 
this is known to be a poor assumption, both from 
direct measurement of the dependence of mass-to-light ratio $M/L$ on luminosity $L$, as manifested by the fundamental plane, and from 
theoretical arguments which suggest that dwarf 
galaxies form stars inefficiently. 
Feedback from star 
formation can partially suppress dwarf galaxy 
formation and thereby flatten the slope of the 
resulting luminosity function.  The apparent increase in the 
comoving  number density of dwarfs with increasing redshift may 
be a manifestation of such a process. 

 Accounting for 
the characteristic luminosity $ L_{\ast}$  of bright 
galaxies presents a more fundamental problem.  In 
hierarchical clustering, there is no limit on the mass 
accumulated, other than that set by the age of the 
universe.  Yet galaxies are clearly distinct in 
morphology and luminosity density from galaxy 
clusters.  Bright galaxies have a characteristic 
luminosity, defined by the Schechter luminosity 
function, $ L_{\ast} \approx 10^{10} h^{-2} L_{\odot}. $
Explanations of $ L_{\ast} $ have hitherto been based 
on the requirement that baryonic matter must cool 
within a specified time-scale in order to form stars 
with even moderate efficiency.  However this 
constraint does not restrict $L_{\ast} $ to lie in the
range of galaxy luminosities, for objects of cluster mass are forming at 
present and the cooling time in rich cluster cores is 
generally less than a Hubble time.

I  review the cooling constraints on galaxy formation (Section 4),
and then consider star formation in spheroidal protogalaxies.
In Section 5, I develop feedback arguments that arise from the impact of massive star
formation and death on protogalaxies. Massive galaxies must form stars efficiently. The protogalactic environment must  therefore
both  be able to cool efficiently to form stars, and yet maintain radiative balance with energy input from dying stars. One can thereby
 account for the characteristic luminosity of a galaxy. On the other hand, in dwarf and in gas-poor galaxies,
 the ejecta from supernovae drive galactic winds via the porosity of the 
volume-dominating hot phase (Section 6). A final section summarizes these various results and their implications.

\bs\goodbreak

\noi\c{\bf 2. Disk Star Formation}
\medskip
The theory of large-scale gravitational instability successfully accounts for many aspects of star formation in the Milky Way and in nearby disks. 
I will argue that there are  two key parameters:
the interstellar medium porosity $P$, which is a measure of the supernova remnant-heated volume fraction, and the Toomre parameter $Q$, which controls disk stability. It is the interplay between $P$ and $Q$ that provides the necessary feedback that allows disks to be long-lived,
and, I shall hypothesize, self-regulated.

Simple global star formation models can account for many aspects 
of disk star formation. These include the star formation rate, metallicity and gas surface density  as a function of disk radius and age, as well as the metallicity distribution of disk stars (Prantzos and Aubert 1995).
These models are   based on a semi-phenomenological treatment of disk stability
(Wang and Silk 1993), and are generally
confirmed by numerical simulations (Steinmetz and Muller 1995).
Key observational motivations include the near constancy of the star formation rate over galactic disk age, the proportionality of star formation rate to  gas surface density $(HI + H_2)$,
the fact that the inner regions of star-forming disks are marginally unstable $(i.e. \ \ Q\sim 1)$, and that below a surface density threshold defined by $Q \simgt 1 $,
global disk star formation effectively ceases, at least in giant HII regions (Kennicutt 1989).
Common to these models is a dependence of star formation rate on differential rotation rate, which controls both the  linear growth of gravitational
instabilities in the disk and the coalescence rate of molecular clouds
(Wyse and Silk 1989).

Define an efficiency of star formation per dynamical 
time by writing
$$ \epsilon = \dot \rho_{\ast} t_{dyn} / \rho_{gas}, 
\eqno\neweq\eqnam{\eff}$$
where $\rho_{gas}$ is the gas density, the star formation rate 
$\dot \rho_{\ast}$ can be written as 
$\dot\rho_{\ast}=r_{SN}m_{SN},$ $ r_{SN}$ is the supernova rate per unit volume,
and  $m_{SN}$ is the mass 
in stars formed per Type II supernova. 
For a solar neighborhood (or Miller-Scalo) initial mass function, 
$ m_{SN}\approx 250 M_\odot $. 
One can write
the generic disk star formation rate in the form, from \eff ,
either locally by $\dot \rho_{\ast}=\epsilon\rho_{gas}t_{dyn}^{-1},$ 
or globally as  $\dot M_{\ast}=\epsilon M_{gas}t_{dyn}^{-1},$ (since gas and young stars have similar radial distributions and scale-heights).
Disks are observed to
have low efficiency at forming stars, $\epsilon\sim 0.03,$
and this is of course required to maintain the gas supply needed for ongoing star formation.

Now the porosity is given by
$$P=\nu_{SN} \dot\rho_{\ast}m_{SN}^{-
1}=\epsilon{\nu_{SN}\over m_{SN}}{\rho_{gas}\over 
t_{dyn}},
\eqno\neweq\eqnam{\por}$$
where the SNR 4-volume in the SNR cooling phase is (Cioffi, Mckee and Bertschinger 1988)
$$  \nu _{SN} = 7.82 \times 10^{12} p^{-1.36}_4 n^{-
0.11} \zeta^{-0.2}E_{51}^{1.27}{\rm  pc^3 yr} \equiv A p^{-1.36}_{gas} 
\rho^{-0.11}_{gas} \zeta^{-0.2}E_{SN}^{1.27}, 
\eqno\neweq\eqnam{\nusn}$$
$\zeta$ is the gas metallicity relative to the solar value,
and the gas pressure $p_4\equiv 10^4(p_{gas}/k)\, \rm cm^{-3}K.$
I take $  p_{gas} = \rho_{gas} \sigma_{gas}^2 ,$
where $\sigma_{gas}$ is the gas velocity dispersion.
Note that supernovae are a stabilizing influence: as the pressure increases, the porosity is reduced.

I will argue that the star formation efficiency $ 
\epsilon $ is determined by requiring the porosity $ 
P $ of the interstellar medium to not be large, and 
thereby avoid blow-out.  I consider the local  star formation efficiency
 $ \epsilon $, defined by \eff\ .
It is plausible to believe 
that self-regulation must result in maintaining $ P 
\sim 1 $, since blow-out $ ( P \gg 1) $ reduces the 
(massive) star formation rate, while $  P \ll 1 $ 
would allow the cold phase to dominate sufficiently 
that the star formation rate increases. Indeed for our own interstellar medium, observations show that $ P \sim \cal O$$ (1), $ although there are contentious arguments about 
whether $ P = 0.2$  or 1  is closer to what is seen in the solar neighborhood
(Shelton and Cox 1994).  
By adopting the 4-volume \nusn\
 swept out by a supernova 
remnant that terminates its expansion at ambient gas 
pressure $ p_{gas} $,
 I infer from \por\ and \nusn\ that
$$ \epsilon = {(P / A)} m_{SN}\zeta^{0.2}E_{SN}^{-1.27} \rho^{0.47}_{gas} \sigma 
^{2.72} _{gas} t_{dyn}. \eqno\neweq\eqnam{\efsfr}$$

Now I suppose that interstellar clouds are accelerated 
by supernova remnants and decelerated by cloud 
collisions.  The clouds must acquire a terminal 
velocity given by
$ \sigma _{gas} =  \epsilon v_{SN}{(t_{coll} / t_{dyn})}, 
$
where  the   specific momentum injected by supernovae per unit of gas mass that forms stars   is defined by 
$$v_{SN} = { E_{SN}  v_c^{-1} m_{SN}^{-1}}= 500E_{51}^{13\over 14}n_{gas}^{-{1\over 7}} m_{250}^{-1}\zeta^{-3/14}\, \rm km\, s^{-1}.$$
Here  $v_c\approx 413\, E_{51}^{1\over 14}n_{gas}^{{1\over 7}}\zeta^{3/14}\rm \, km\, s^{-1}$ is the velocity at which an expanding remnant first undergoes substantial radiative losses. 
The 
cloud collision time-scale can be written
$ t_{coll} = 2^{1/2}(\mu _{cl} / \mu _{gas}) (H /
\sigma _{gas} ). 
$
Here,  $ \mu _{cl} $ is the cloud column density, $ \mu _{gas} 
$ is the column density of gas in the galactic disk, 
and  $H$ is the scale-height of the disk.  If the clouds are 
marginally bound, as seems to be the case in our 
interstellar medium, and maintained against 
gravitational collapse by internal pressure support
that is approximately equal to the mean interstellar 
pressure, one has
$ \mu _{cl} = (p_{gas} / G) ^{1/2}. 
$
The scale-height
 $ H =\mu _{gas}/\rho _{gas}$,
and so $ t_{coll} = (6\pi G \rho _{gas})^{-1/2}.$
 Combining  these expressions,
one finds that
$$ \sigma_{gas} = 
6.90 P^{-0.58}n_{gas}^{0.1}\, E_{51}^{0.2}\zeta^{0.008}\rm km\, s^{-1}. \eqno\neweq\eqnam{\sigd}$$

Hence porosity self-regulation suffices for the gas to have 
constant velocity dispersion. As $P$ increases, momentum transfer is progressively less efficient, and $\sigma_{gas}$ decreases. 
Only about 2-3 percent of the injected supernova energy
is expended  
in supplying momentum to the interstellar gas.
There is no dependence of disk velocity dispersion on the IMF.
{\it It is remarkable that the gas velocity dispersion is insensitive to all physical parameters, and for a self-regulated $(P\sim 0.5)$ disk is close to the observed value, }  of about $11\, \rm km\, s^{-1}$ for the 3-dimensional 
peculiar velocity dispersion of interstellar molecular clouds  within 3 kpc of the sun (Stark and Brand 1989).

Presumably, the young stars which dominate disk light  have the same velocity dispersion and scale-height as the gas.
One implication is that the (thin) disk scale-height ($\equiv \sigma_{gas}^2/G\mu$; $\mu$
is disk surface density)
is constant, as observed for the stellar component  both in edge-on thin disks
as well as  for the thick ({\it i.e.} older)  components (de Grijs and van der Kruit 1996), only provided that disk surface density is constant. The disk surface density primarily comes from stars, so  I conclude that Freeman's law of constant central surface brightness for luminous
spiral galaxy disks is equivalent to the requirement of constant  
scale-height.  Another implication is that low surface brightness galaxies are expected to have thicker disks than normal galaxies.
The predicted constancy of  disk gas velocity dispersion is likely to be  the driver behind  both  disk scale-height and  surface brightness in a more
realistic model of disk evolution, 
because the stars form from the gas, but this issue is beyond the scope of the present discussion.

Next, consider the star formation rate. First I evaluate the star formation efficiency. Now $t_{dyn}=(2\pi G \rho H/R)^{-1/2},$ so that 
$$\epsilon ={(\sigma_{gas} /v_{SN})}(\rho_{gas}/ \rho)^{1/2}
(3/ \pi)^{1/2}.\eqno\neweq\eqnam{\epsd}$$
Note that $\epsilon$ decreases with time as the gas fraction decreases.
One can  write the star formation rate, using  \efsfr , as 
$$\dot\rho_\ast=\rho_{gas}^{1.74}P^{-0.58}\alpha,\eqno\neweq\eqnam{\sfrv}$$
where
$$\alpha=A^{0.58}m_{SN}\zeta^{0.22}E_{SN}^{-0.73}v_0^{1.58}(6\pi G)^{0.79}
=10.29m_{250}.$$ 
In the absence of accretion, one finds the solution
$\rho_{gas}=\rho_i(1+t/t_\ast)^{-1.35},$
where the characteristic star formation time-scale is
$$t_\ast=(2/3)P^{0.58}\alpha^{-1}\rho_i^{-0.74}= 0.81\, P^{0.58}n_i^{-0.74}m_{250}^{-1}E_{51}^{0.73}\zeta^{-0.22}\,\rm Gyr.
\eqno\neweq\eqnam{\sft}$$
The return of mass from evolved
stars is easily incorporated as a correction factor into this and other expressions given here for star formation times and rates.

These results have two noteworthy implications. At constant scale-height, the star formation rate per unit disk surface area is proportional to the gas surface density  to the 1.74 power, with a  dependence on just  one
 parameter: porosity. Not only is this a reasonable fit to data on $H\alpha$ surface brightness (Kennicutt 1989), but there is a straightforward  prediction that at a given gas column density, $H\alpha$ surface brightness is proportional to $P^{-0.58}.$ Moreover, rapid star formation is achieved in systems with high initial gas density. This has obvious implications for star formation in early type, bulge-dominated galaxies, where the past
star formation rate is inferred to be high (Kennicutt, Tamblyn and Congdon 1994).

I now derive   the global star formation rate.
Knowledge of the star formation efficiency $\epsilon$
is the key.  
Integrating the star formation rate per unit volume \sfrv\ over
disk volume, and making use of $R=2^{1/2}t_{dyn}v_{rot},$
 yields
$$\dot M_\ast=6.74P^{-0.58}v_{rot,200}^3n_{gas}^{0.24}
\left({{\rho_{gas}/\rho}\over 0.1}\right)^{3/2}
\left({{R/H}\over 0.1}\right)^{1/2}\,\rm M_\odot\, yr^{-1},\eqno\neweq\eqnam{\sfrd}$$
where $v_{rot,200}\equiv v_{rot}/200\,\rm km\, s^{-1}.$
To proceed further, it is necessary to decide  on the physics that controls the disk gas fraction.

In fact, so far, disk self-gravity has not been utilized.
The key  to determining  $\rho_{gas}/\rho$ is via consideration of  dynamical self-regulation.  Define the Toomre parameter by
$$ \tilde Q = {{\Omega \sigma _g}/ { (\pi G \mu _{gas}  \beta)}};\ \ \ Q=\tilde Q \beta
\eqno\neweq\eqnam{\toom}$$
appropriate for a flat rotation curve, where $ \beta = 1 
+ { (\sigma _g / \sigma_\ast)(\mu _{\ast}  \mu _{gas} )} $
approximately corrects for the self-gravity of the 
stellar component (velocity dispersion $\sigma_\ast$, surface density $\mu _{\ast}$) 
and   $\Omega$ is the disk angular velocity.  One can also express $Q$ as $ \mu _{cr} / \mu _{gas},$ where $\mu _{cr} \equiv 
\Omega \sigma _g / \pi G . $  One finds 
empirically for spiral disks that $ Q \sim 1 $ 
throughout the star-forming region.  Using Q for the 
moment as an independent variable, one can write
 the local
gas fraction as
$$\rho_{gas}/\rho =0.017 n_{gas}^{0.1} P^{-0.58} Q^{-1}v_{rot,200}^{-1}E_{51}^{0.2}\zeta^{0.008}\delta,
\eqno\neweq\eqnam{\fg}$$
where $\delta$ is the ratio of disk to gas scale-heights.
If $P$ and $Q$ self-regulate with  $P \sim Q\sim 1$,
 I have inferred  the 
 gas fraction.

One now has $$ \epsilon   
 =0.07 n_{gas}^{0.19}m_{250} P^{-0.29}Q^{-1/2}v_{rot,200}^{1/2}E_{51}^{-0.83}\zeta^{0.22}\delta^{1/2}. 
\eqno\neweq\eqnam{\epsb}$$  
Inserting the expression \fg\ for the 
gas fraction into equation 
\sfrd,
 one finds that the star 
formation rate is 
$$  \dot M_{\ast} 
=1.4 v_{rot,200}^{5/2}n_{gas}^{0.29}m_{250} P^{-0.87}Q^{-3/2}E_{51}^{-0.63}\zeta^{0.22}\delta^{3/2}\,\rm M_\odot\, yr^{-1}. \eqno\neweq\eqnam{\sfrfin}$$
The inferred star formation rate for the Milky Way Galaxy, with $P\approx 0.3$, $Q\approx 1,$ $\delta\approx 1$
and $v_{rot}= 220 \,\rm km\,s^{-1},$ is about $5\,\rm M_\odot\, yr^{-1},$
in good agreement with the observed value ({\it e.g.}  McKee 1989; Noh and Scalo 1990).
Since  the disk becomes more unstable as $Q$ decreases, the star formation rate must increase and the ensuing massive star formation will drive up the porosity $P$, which in turn must have the effect of reducing the cold gas supply and thereby depress the star formation rate. Hence this expression for the star formation rate provides an explicit demonstration of disk self-regulation.
Moreover, the self-regulation implies that the associated dispersion in    
$\dot M_{\ast}$ as a function of $v_{rot}$ will remain small.

\bigskip\goodbreak
\noi\c{\bf 3. The  Tully-Fisher Relation}
\medskip

Dark matter is  irrelevant to the derivation of the star 
formation rate \sfrfin . Gas disk self-gravity includes a contribution from the stars, but dark matter plays  a subdominant role in maintaining the rotational velocity in the luminous disk region,
as is observed for optical rotation curves
(Kent 1988).  Even in the outer parts of disks, the stellar component
 is usually close to its maximum possible value and  the shapes of the luminosity profiles and HI rotation
curves are correlated, while the relative contributions of the halo
and stellar components to the rotation velocity vary significantly
with luminosity and/or morphological type  (Kent 1987).
While low surface brightness and dwarf galaxies are usually dark matter-dominated ({\it e.g.} Cote, Carignan and  Sancisi  1991), even here there  are notable counterexamples
({\it e.g.} Carignan, Sancisi and van Albada 1988).

The preceding result \sfrfin\ is equivalent to  a derivation of the  Tully-Fisher relation in the blue band.
The blue Tully-Fisher relation is dominated by 
light associated with current star formation, and the 
self-regulation of disks $ (Q \sim 1) $ therefore 
predicts a slope $ \alpha\approx 2.5$ where $L \propto 
v_{rot} ^{\alpha}.$  This slope is close to what is observed in the $B$ band.
The low dispersion in the  Tully-Fisher relation may perhaps be understood in terms of $P$ and $Q$ self-regulation. Of course, dark matter, and its cosmological evolution, is necessary to establish the actual range of observed rotational velocity and initial gas disk mass. It is the transformation to luminosity that is driven by self-regulation.

In fact,  compilations of 
Tully-Fisher data for available samples (Burstein et al. 1995; Strauss and Willick 1995) find that 
the slope increases systematically with increasing 
wavelength: $ \alpha = 2.1-2.2 \ (B), \ 2.5  \ (R), \ 2.7 \ (I) $ and 
$4.1  \ (H).$  A recent comprehensive $I$ band analysis of a large sample of galaxies  finds 
$ \alpha = 3.1$ (Giovanelli {\it et al.} 1997). In the $I$ band,
and especially in the
$H$ band, one is measuring  the old stellar populations and 
therefore needs to include the dominant 
contribution from stars formed over the entire history 
of the disk. 

The  old stellar populations may be responsible for  the
observed steepening of the Tully-Fisher relation.
For example, Dopita and Ryder (1994) find that 
$\dot\mu_\ast\propto\mu_I^{0.64},$
which would result in  prediction of  steeper $I$, and presumably $H$
if a similar relation extends to longer wavelengths, band slopes relative to the $B$ band slope. For example, if naively applied to the observed blue
slope,  this observed correlation would steepen the Tully-Fisher slope from 2.1 to 3.3. The steepening  is less if not all of the $B$ light is associated with current star formation.
Hence an explanation of the blue Tully-Fisher relation  seems to account 
for the Tully-Fisher relation at longer wavelengths. This suggests that the concern (Willick 1996) that  most of the observed steepening in the $H$ band may be due to use of aperture magnitudes rather than total magnitudes, as used in the other bands where the entire galaxy is imaged, may not be valid.

Low surface brightness (LSB) galaxies present a challenge to any explanation of the Tully-Fisher relation. These galaxies follow the same Tully-Fisher relation
as do normal galaxies, at least in the $B$ band (Zwaan {\it et al.} 1995),
so that application of the  virial theorem implies that if $L/v_{rot,max}^4=constant,$ then $(M/L)^2\mu_\ast=constant.$ The LSB galaxies, typically a factor 4 lower in central surface brightness than normal galaxies in the sample studied by  Zwaan {\it et al.}, are then inferred to  have twice the mass-to-light ratio of normal galaxies,
 and hence  are also a factor of 2 larger in disk scale-length  at given $L$ and $v_{rot,max}, $ as observed. In practice, the observed $B$ band Tully-Fisher relation has a slope that differs from 4: if I take the observed slope of, say, 2.2, and apply the virial theorem, I deduce that  
$(M/L)^2\mu_\ast \propto L^{0.8}$  At fixed luminosity, the previous conclusion about the increase  in $M/L$ for the LSB galaxies still applies.

However $Q$  should increase,
approximately inversely with $\mu_{gas}$, for LSB galaxies. This helps explain why the LSB galaxies form stars  per unit surface area at a lower rate per unit mass of gas than do normal galaxies. 
The characteristic star formation time \sft\ is long  because the initial gas density is low. As disk stability,
characterized by $Q$,  increases, I expect that the star formation rate per unit area and the porosity
$P$ must decrease. Perhaps the gas fraction that forms stars per unit
 dynamical time is determined by local cloud properties and is constant: one would then infer that $\epsilon$ is constant and therefore that $QP^{0.6}=constant.$ In this case, 
the Tully-Fisher relation \sfrfin\ derived above for normal galaxies is identical for LSB galaxies.  One does not need to appeal to the
dark matter distribution or to the initial specific angular momentum of the protogalactic precursors to resolve the question of why the LSB galaxies satisfy a normal galaxy Tully-Fisher relation. Of course, these other issues must presumably be invoked to explain why LSB galaxies have higher $M/L$ ratios  and/or larger scale-lengths than normal galaxies. However the Tully-Fisher relation
is entirely a matter of disk star formation physics, which provides a mechanism for self-regulation
of the gas reservoir.
\bs
\goodbreak
\noi\c{ \bf 4. Cooling Constraints}
\medskip

I turn now to the question of what determines the characteristic luminosity
of a spheroid-dominated galaxy. Cooling is generally considered to be the key to understanding the luminous mass of a galaxy (Rees and Ostriker 1977; Silk 1977).  However cooling does not necessarily lead to galaxy formation.  Cooling flows in cluster cores are environments where one might expect to see forming galaxies. Only old galaxies are found in cluster cores. 
Even if cooling flows were to have formed giant cD galaxies in the past,  
one has to  remember that   presumed hosts of past, as well as  current, cooling flows, namely  many 
clusters and groups, do not contain dominant cD's. 

 Theoretical arguments converge to a similar conclusion. Specifically,
one can straightforwardly show 
that the mass of gas  within a dark matter 
potential well that can cool within a Hubble time is 
limited only by the mass of dark matter, and therefore cannot account for the luminous stellar mass. 
Consider the 
collapse of gas within a dark halo, represented by an isothermal sphere of cold dark matter that 
contains gas fraction $ f_{gas} $ with density $ \rho = 
\sigma ^2 / 2 \pi G r^2 $, constant velocity 
dispersion $ \sigma $, and mass $ M (<r) = 2 r \sigma 
^2 /G. $    In massive halos,
$ T 
\approx {\sigma ^2  m_p / 3k } = 4 \times 10^6{\rm  K}
 \left( {  \sigma / 300{\rm km  \, s^{-1} }  } \right)^2  .$
The ratio of gas cooling time at radius r 
to Hubble time is 
$$ { t_{cool} \over t_H } = {3 nkT \over \Lambda n^2 
t_H } = { m^2_{p} 2 \pi G r^2 
\over f_{gas}\Lambda  t_{H} } \equiv \left( {r \over r_{c,H}}\right)^2, 
\eqno\neweq\eqnam{\tc}$$
where the cooling radius 
$$ r_{c,H} = ( f_{gas}\Lambda  t_H / 2 \pi G )^{1/2} m^{-
1}_{p} =0.3(f_{0.1}\Lambda_{24} t_{15})^{1/2}\, \rm Mpc
\eqno\neweq\eqnam{\rc}$$
and the cooled mass
$$ M ( <r_{c,H} ) = { \sigma ^2  G^{-3/2} m_{p}^{-1} } 
( 2 f_{gas}\Lambda  t_H / \pi )^{1/2}
=10^{12}\sigma_{100}^2(f_{0.1}\Lambda_{24} t_{15})^{1/2} \,  M_\odot. 
\eqno\neweq\eqnam{\mc}$$
Here $\Lambda_{24}\equiv \Lambda/10^{-24}\rm \,erg\, cm^3\,s$ is the cooling rate, $t_{15}\equiv t/{15\,\rm Gyr}$ is the age of the galaxy, 
 $f_{0.1}\equiv f_{gas}/0.1,$ and $\sigma_{100}\equiv \sigma/\rm 100 km  \, s^{-1}.$ 

Gas cooling within a Hubble time might be relevant to 
disk galaxy masses, which accumulate by slow infall.  
One might also expect star formation to occur 
efficiently within a dynamical time, $ t_{dyn} = 
r/\sigma $, as has been argued for elliptical galaxy 
formation.  In this case, 
$${ t_{cool} / t_{dyn}} 
= { 2 \pi rm_p^2 \sigma  f_{gas}^{-1} \Lambda^{-1} } \equiv { r 
/ r_{c,d} }, 
\ \ {\rm or} \ \ 
r_{c,d} = 
0.3 f_{0.1}\Lambda_{24} \sigma_{100}^{-1} \rm Mpc. 
\eqno\neweq\eqnam{\rcd}$$  The mass that has cooled within a 
dynamical time is $$ M ( <r_{c,d} ) = { \Lambda \sigma 
f_{gas} \over \pi G^2m_p^2 }
=10^{12}f_{0.1}\Lambda_{24} \sigma_{100} M_\odot. 
\eqno\neweq\eqnam{\mcd}$$ 

 In the relevant temperature range ($T\simgt 
10^7\rm K$), one can write
 $\Lambda \approx \Lambda_{ff} = 
2 \times 10^{-27} T^{1/2} \rm erg \, cm^3 s^{-1} 
\equiv 
\Lambda _0 \sigma,$ 
 and 
this expression is  appropriate at $T\simgt 
10^6\rm K$ if the metallicity is very 
low.  In this case,  
$$ M (<r_{c,H}) = 
{\left( 2 f_{gas} \Lambda _0 t_H \over \pi \right)}^{1/2} { \sigma ^{5/2} \over G^{3/2} m_{p} 
} =10^{12}\sigma_{100}^{5/2}f_{0.1}t_{15}^{1/2}  M_\odot
\eqno\neweq\eqnam{\mffh}$$
and 
$$M (<r_{c,d}) = ({ \Lambda _0 f_{gas} \over \pi G^2 
m_p^2 } ) \sigma ^2
=10^{12}\sigma_{100}^{5/2}f_{0.1} M_\odot. 
\eqno\neweq\eqnam{\mffd}$$   Both mass estimates increase 
without limit as the galaxy halo potential grows, as also found in simulations by Thoul and 
Weinberg (1995).

Cooling  and feed-back constraints have been incorporated into hierarchical galaxy formation using semi-analytic models. However the sharp decline in the galaxy luminosity function
above $L_\ast$ is not explained. For example, Kauffman, White and
Guiderdoni (1993) introduced an arbitrary cut-off to avoid formation of excessively luminous galaxies. Dekel and Silk (1986) demonstrated that feedback helps suppress formation of dwarf galaxies, and later papers incorporated this effect into   hierarchical galaxy formation  (Lacey and Silk 1991; Lacey {\it et al.} 1993; Kauffmann, Guiderdoni and White
1994; Cole  {\it et al.} 1994).
I now argue that combining the physics of cooling and feedback helps suppress the formation of overly massive galaxies.

\bs
\goodbreak
\noi\c{ \bf 5. A Derivation of $L_\ast$}
\medskip

I  consider supernova heating and feedback, as a 
possible means of limiting the mass of cooled gas.  
Suppose supernovae occur at rate $ R_{SN} $ and each 
supernova injects $ E_{SN} $ ergs into the 
interstellar gas, of total mass $ M_{gas}$.  For thermal 
balance to occur, the gas must be able to radiate away 
the energy injected.  The specific rate of thermal 
energy radiated is $ { 1 \over 2} \sigma ^2 t_{cool}^{-
1} $, and this is therefore set equal to the 
injected energy rate $ R_{SN} E_{SN} / M_{gas}. $  
I will argue below that this situation is stable for massive protogalaxies, and does not lead to a supernova-driven wind.
From the generic expression for star formation
efficiency \eff , I  then obtain
$$\sigma^2=\left(t_{cool}\over t_{dyn}\right)2\epsilon 
{E_{SN}\over m_{SN}},
\ \ {\rm or} \ \ \sigma=270\left(\epsilon_{0.2} 
E_{51} m_{250}^{-1}({t_{cool}\over t_{dyn}})\right)^{1/2}\rm km \, s^{-1}.
\eqno\neweq\eqnam{\sig}$$
Note that the  star formation
efficiency ($\epsilon\equiv 0.2 \epsilon_{0.2}$) is expected to be about 10--20
percent for protoellipticals, as inferred from 
population synthesis modelling of nearby and distant 
galaxies, which requires most of the stars to have 
formed within 1-2 Gyr (Bruzual and Charlot 1993).
A lower efficiency would be difficult to reconcile with starbursts.
 One would certainly need
$\epsilon > 0.3$ in order to have most star formation underway by $\sim 30 t_{dyn}$, or $\sim 3$ Gyr.

The value
$ m_{SN}\approx 250 M_\odot $ is based on 
scaling to the Milky Way, where the specific SN I rate, which must also be included in momentum injection considerations,  is 
higher than in young galaxies and the star formation rate is $ \sim 5 M_\odot / 
\rm yr. $    However for an IMF enriched 
in massive stars, $ m_{SN}$ may be substantially 
smaller.  For example, the most extreme possibility 
considered is that  the cluster metallicity, 
including intracluster gas, is a monitor of the 
protoelliptical yield, from which one infers that the 
yield is approximately 4 times higher (in terms of 
mass of iron per unit mass of stars) than in the Milky 
Way (Renzini {\it et al.} 1993).  One could achieve this high a yield by lowering 
$ m_{SN}$ by a corresponding factor, to a first 
approximation (Elbaz, Arnaud and Vangioni-Flam 1995).

To form bulges and ellipticals, not only must thermal balance be attained, but efficient star formation is required.
Population synthesis modelling for both ellipticals and bulges suggests that the characteristic star formation time $t_\ast$ is less than a Gyr. Theoretical arguments require the star formation time to not exceed the dynamical time, based on  
diverse considerations that include  dynamical friction settling in major mergers, cloud coalescence
and cloud disruption by massive star formation (Silk and Wyse 1997).
In order to form stars efficiently, one certainly requires $ t_\ast< t_{cool} < t_{dyn} $ : 
this is essential in order to first produce the supernovae that heat the 
gas, otherwise the gas is too hot to form stars.  Hence the conditions to form a galaxy are thermal balance and $t_{cool}< t_{dyn}.$ The 
requirement of thermal balance for the 
protogalactic gas now leads to an upper limit on velocity 
dispersion:
$ \sigma \simlt \sigma_\ast\equiv 270 ( \epsilon_{0.2} E_{51} 
  m_{250}^{-1} )^{1/2} \rm \, km\, s^{-1}.$ 
 The central velocity dispersion of an $ 
L_{\ast} $ elliptical galaxy is approximately $ 270 
\, \rm km\, s^{-1},$ and is obtained as a limiting value if $\epsilon \approx 0.2.$

In other words,  thermal support of the gas sets a limit on $ 
\sigma $, and therefore on galaxy luminosity, since $L$ is 
correlated with $ \sigma $ according to the Faber-Jackson relation.  It is interesting to note that a 
byproduct of the supernovae, metallicity, correlates 
more tightly with $ \sigma $, and in particular with 
local $\sigma$, than with $L$  (Fisher, Franx and Illingworth 1995).
This suggests that the 
potential well depth, characterized by $ \sigma $, is 
more fundamental to early star formation in 
ellipticals than the total mass in stars.  The 
explanation for a critical luminosity $ L_{\ast} \approx 
10^{10} h^{-2} L_{\odot} $ above which the number of 
galaxies exponentially declines may therefore lie in 
the upper limit on $ \sigma $ for a protospheroid.  
Note that spheroids dominate at $ L \ge L_{\ast} $: 
the galaxies with largest $ \sigma $ (and $L$) are 
giant ellipticals and  early-type (bulge-dominated) 
disk galaxies. 

For disks and ellipticals to  have similar potential well depths, one
must have $\epsilon (t_{cool} / t_{dyn})\approx constant.$
Supernova momentum input into the interstellar medium has an efficiency of a few percent, and the dissipation time of the gas  can be as long as the age of the disk: indeed a long timescale is inevitable from the simple observation that star-forming disks are gas-rich. One may regard the cooling time as a lower bound on the star formation time. Perhaps, if $t_{cool}$ is interpreted more loosely, one could take the momentum dissipation  time-scale as an actual estimate of the star formation time.
Hence one would end up with 
$ \sigma_\ast\propto ( \epsilon t_\ast/t_{dyn})^{1/2} ,$ and therefore
similar values for $\sigma_\ast,$ both in 
protoellipticals that form stars with high efficiency over a dynamical time 
and in protodisks that form stars with low efficiency over many dynamical times.
One can make a reasonably strong case for self-regulation to have occurred
in disks, and at least in this case actually derive the star formation efficiency \epsb .
\bs\goodbreak
\noi\c{ \bf 6. Protogalactic Winds}
\medskip

I finally show that  for low $ \sigma $ galaxies, thermal balance is unattainable and a protogalactic wind is inevitable. More generally, this occurs even in more massive galaxies 
once the initial gas fraction has dropped below $ \sim 
10$  percent.   A critical parameter in ascertaining 
the viability of a galactic wind is the porosity of 
the interstellar gas that is determined by the network 
of expanding and interacting supernova remnants.  
 Radiative losses regulate the wind velocity, and it is  reasonable to estimate the effective wind velocity as given by the specific momentum injected by supernovae. 
If 
the volume fraction $ f $ of hot bubble interiors 
dominates, so that the porosity $ P \gg 1$ (recall 
that $f = 1- e^{-P} ) $, a wind is inevitable provided 
that a momentum balance condition is also satisfied (e.g. Doane and Mathews 1993), namely that the momentum per unit mass of gas that is injected into the interstellar medium by supernovae exceeds the escape velocity, or 
$v_{SN} > \sigma.$ 
This leads to a remarkable coincidence, given the estimated value of $v_{SN}$ : winds can occur (but admittedly do not necessarily occur) in potential wells corresponding to those of depth less than or comparable to $L_\ast$ galaxies.
 A wind is only inevitable in galaxy potential wells with velocity dispersion $\sigma\simlt 500 \rm km\, s^{-1}$ provided also that $P\gg 1.$
I now apply \sig\ to eliminate $ \epsilon/m_{SN} $ from \por, and use \nusn\ to obtain
$$ P = \sigma ^{-
0.71}\left(\rho\over\rho_{gas}\right)^{0.5}\rho_{gas}^{0.04}  ({ A 
\over 2}E_{SN}^{0.27} \zeta^{-0.2}) ({t_{dyn} \over t_{cool} }) (\rho^{0.5}t_{dyn})^{-1}.\eqno\neweq\eqnam{\porf}$$
Note that for a top-heavy initial mass function, $m_{SN}$ would decrease. However at fixed velocity dispersion $\sigma$, $\epsilon/m_{SN}$ is constant, so that the porosity $P$ is independent of the IMF.

Cooling certainly permits star formation if $ t_{cool}> t_{dyn}$ as long as 
$ t_{cool}< t_{0}$, the present age of the universe. This may be the relevant condition in disk galaxies, where the star formation efficiency is at most a few percent.
However in order for stars to form much more efficiently, one certainly requires
$ t_{cool}< t_{dyn}$. This condition is probably essential for protoelliptical formation as well as in starbursts.
Inserting numerical values for the various constants,
I find that  
$$
P> 0.64\sigma_{100}^{-0.71}
f_{0.1}^{-0.5} n_{gas}^{0.04}E_{51}^{0.27}\zeta^{-0.2}\gamma^{-1},
$$
where 
$\gamma\equiv(2\pi G\rho)^{0.5}t_{cool} \approx t_{cool}/ t_{dyn} \simlt 1.$
Within the starburst core, the porosity is likely to be large and constant
if $\sigma$ and $f_{gas}$ are sufficiently small.

Clearly, one cannot avoid  high porosity either at  low gas velocity dispersion or gas fraction, or in the core. The conditions for a radiatively unstable wind are satisfied. To form stars, cooling is essential. Hence protogalactic winds must undergo radiative cooling and consequently be unsteady. 
 I infer that a wind is inevitable either at low $\sigma$ or low 
$ \rho_{gas} / \rho $, independently of 
star formation efficiency. At $\sigma \simgt 100f_{0.1}^{-0.7}\rm km\, s^{-1},$
a wind is inhibited, since $P<1.$ However the earlier discussion requires
$\sigma \simlt \sigma_\ast$ in order for the supernova energy input to be radiated, otherwise star formation is suppressed. This helps one understand why luminous galaxies have a relatively narrow range of $\sigma .$ 

\bigskip

\goodbreak

\noi\c{ \bf 7. Discussion}
\medskip
Global star formation can be understood via self-regulation. 
This involves feedback from star formation. In this paper, I have 
developed feedback arguments that arise from the impact of massive star
formation and death on gas-rich galaxies and protogalaxies.
Star formation in disks has a  more secure foundation in theory and in phenomenology than star formation in spheroids. 

The idea underlying Section 2 is that in a self-gravitating gas disk, nonaxisymmetric gravitational instabilities drive cloud coagulation, collapse and star formation. Supernova explosions, as well as HII regions, stir up the interstellar gas, tending to increase the gas velocity dispersion and the gas 
scale-height. The feedback operates via the overlapping hot interiors of supernova remnants that accelerate swept-up shells of interstellar matter and drive gas out of the disk via interstellar chimneys,  ultimately generating a galactic wind if the volume filling factor of the hot gas is sufficiently high. 
The observed self-regulation of star formation in disks is effectively controlled by two dimensionless parameters: the Toomre gravitational instability parameter $Q$ and the porosity $P$ of the interstellar medium to supernova remnant-heated gas.
I have argued that $P$ and $Q$ act in concert: as
$Q$ decreases, $P$ increases, with the consequence that the star formation rate, found at specified rotation velocity to reduce to a simple function of $P$ and $Q$, self-regulates.  The interplay between $Q$ and $P$ leads to an
explanation of the blue band Tully-Fisher relation, which is dominated by ongoing star formation.  

Of course, the rotation curve, as well as the total mass of the disk,  is taken to be specified in this analysis. In fact, the dark matter distribution  must account for the rotational velocity, and the primordial baryon fraction in conjunction with initial conditions accounts for the total cooled mass of stars and gas (Navarro and Steinmetz 1997). However the Tully-Fisher relation, and its low dispersion,  are due to self-regulation of disk star formation.
There is an interesting implication: there is likely to  be a broad dispersion in dark mass, and hence also in gas mass, at fixed rotation velocity or luminosity, if disks self-regulate. Cosmological initial conditions indeed imply a broad dispersion in galaxy masses at specified circular velocity (Eisenstein and Loeb 1996).
The total gas mass is potentially observable, and since the local
gas fraction in the star-forming disk is unchanged, the gas distribution, in the case of additional gas mass,  must be more extended.

Star formation efficiency is high in deep potential 
wells and in gas-rich systems.  In particular, the characteristic time-scale
for star formation is found to be proportional to the inverse 3/4 power of the initial gas density. Given that spheroids have a higher central surface brightness,
and therefore density,  than disks
by about two orders of magnitude, one can understand why early,
spheroid-dominated  Hubble types form stars more efficiently than later,
disk-dominated  Hubble types.
This is consistent with the observed star formation rates for disks of varying Hubble type (Kennicutt, Tamblyn and Congdon 1994). A complementary  argument accounts for the longevity of the gas reservoir against depletion by star formation in low surface brightness galaxies.

The present model of disk star formation is reasonably predictive. $P$ and $Q$ should be anticorrelated
as a function of galactocentric radius. One can try to measure $P$ from $H\alpha$ maps, and $Q$ is especially sensitive to the rotation curve and surface brightness. Porosity tends to oppose star formation, so that later Hubble types, which form stars at a lower rate per unit disk mass than earlier types, 
should have higher porosity.  Little is known about porosity in star-forming disks, and it may be possible, for example by azimuthally averaging
$H\alpha$ maps or $HI$ maps
and appropriate subtraction of stellar continuum, to quantify measures of the porosity of the hot component. At  a given gas surface density, $H\alpha$ surface brightness
should  anticorrelate with porosity.

For spheroids, the situation is necessarily less constrained than for disks. One lacks the analog of the theory of disk instability to motivate description of the star formation rate. The only   resort is to pure phenomenology. Massive spheroidal galaxies must have formed stars efficiently. The protogalactic environment   dissipated thermal energy to form stars, while  maintaining radiative balance with energy input from dying stars.
 If the momentum input from supernovae is not dissipated, or
if cooling is ineffective,  stars do not form.
If it is dissipated too rapidly, more stars form and die until the balance is reestablished.
This conjecture helps  account for the characteristic luminosity of a galaxy.
Specifically, 
if the three-dimensional velocity dispersion of the spheroid exceeds
$\sim 300 \rm km \, s^{-1},$ the baryons do not form stars efficiently.
Moreover the reduced cooling efficiency must result in gas heating that, at the very least, maintains the gas velocity dispersion. Over a Hubble time, however, the gas can cool. The implication is that outer halos and galaxy groups contain a reservoir of cold gas clouds. This is by no means inconsistent with inferences from studies of quasar absorption line systems ({\it e.g.} Steidel, Pettini, Dickinson and Persson 1994; Le Brun,  Bergeron and  Boisse 1996).

 On the other hand, in dwarf and in gas-poor galaxies,
 the ejecta from supernovae drive galactic winds via the porosity of the 
volume-dominating hot phase.
I brought general arguments to bear on momentum input from supernovae that suggest that the luminosities of the spheroidal components are limited by early protogalactic winds. It is inevitable that the porosity  $P\gg 1$ and an early  wind must have
been generated. One would expect to find both gas-poor low surface brightness,
low luminosity dwarf spheroidals and also gas-rich, star-poor  clouds in which star formation has failed to unbind the gas or prevent its later accretion.
Low redshift intergalactic Ly alpha clouds are possible manifestations of such objects (Shull, Stocke and Penton 1996).

Armed with a star formation history, one can speculate about enrichment and chemical evolution.
Formation of spheroids  requires 5-10  times the star formation efficiency that is required to form the disk.
Metallicity is a built-in byproduct, since the supernova rate is inferred once the star formation rate is specified. With a normal IMF, there is  no problem with metal overproduction even when an early wind is generated.
Indeed, the converse applies: there does seem to be a need for enhanced yields from spheroidal systems, as inferred from studies of the intracluster gas in rich clusters of galaxies. The abundance ratios measured for the intracluster gas are characteristic of  Type II supernovae, and  suggest that 
metals were prolifically produced in forming spheroids. 
Most of these metals must have been ejected from the galaxies in early spheroid winds.

Finally, it is tempting to speculate about the origin of the enhanced spheroid yields. For example, 
the observed yields may be  due to a top-heavy IMF in the early phases of spheroid formation, although it certainly is premature to exclude  other factors that can conspire to both enhance star formation efficiency and yields.
If a nonstandard IMF indeed were present,
such early winds can serve the  functions of enriching both the intergalactic gas, as sampled via studies of Lyman alpha absorption systems towards high redshift quasars, and the intracluster gas, where spheroidal systems are inferred, via the correlation of iron mass with light, to be the dominant pregalactic enrichment source. Such winds  simultaneously
allow one to account, at least qualitatively,  for various properties of the stellar components of spheroids that were generated at birth
(Zepf and Silk 1996).
These include enhanced $Mg/Fe$ ratios seen in luminous ellipticals,
the correlation of $Mg$ abundance with local escape velocity in ellipticals,
and even the systematic rise in $M/L$ with spheroid luminosity as encapsulated in the fundamental plane.
Indeed, from the theoretical perspective, such feedback from early spheroid
formation is desirable, if not mandatory, in order to prevent gas overcooling and anomalously small disk formation in hierarchical galaxy formation.
\ms
This research has been supported in part by a grant from NASA. I thank M. Steinmetz and R. Wyse for relevant discussions, and the referee, D. Weinberg, for detailed comments.
\bs
\goodbreak
\noi\c{\bf  REFERENCES}
\def\pp{\parshape 2 0truecm 15truecm .5truecm 14.5truecm}
\def\ref #1;#2;#3;#4{\par\pp #1, {\it #2}, {\bf #3}, #4}
\def\book #1;#2;#3{\par\pp #1, {\it #2}, #3}
\def\rep #1;#2;#3{\par\pp #1, #2, #3}

\def\etal{{\it et al.\ }}
\newcount\refnum

\def\ref{\par\hangindent=1.5em\hangafter=1}
%
\def\aa#1#2{A\&A, {#1}, #2}
\def\aas#1#2{A\&AS\ {#1}, #2}

\def\aj#1#2{AJ { #1}, #2}

\def\apj#1#2{ApJ, { #1}, #2}

\def\mn#1#2{MNRAS, { #1}, #2}

\def\phyr#1#2{Phys. Rep.  { #1}, #2}

\parskip=0cm 
\ref{Bruzual, G. and Charlot, S. 1993, \apj{405}{538}}

\ref{Burstein, D. \etal 1995, in {\sl The Opacity of Spiral Disks} (eds.\ Davies, J. I., Burstein, D.) 73 (Kluwer Academic, Dordrecht).} 

\ref{Carignan, C., Sancisi, R. and van Albada, T.S. 1988, \aj{95}{260}}

\ref{Cioffi, D., Mckee, C. F. and Bertschinger, E. 1988, \apj{334}{252}}

\ref{Cole, S. {\it et al.} 1994, 1993, \mn{271}{781}}

\ref{Cote, S., Carignan, C. and Sancisi, R. 1991, \aj{102}{904; 1231}}

\ref{Doane, J.S. and  Mathews, W.G. 1993, \apj{419}{573}}

\ref{Dopita, M. A. and Ryder, S. D. 1994, \apj{430}{163}}

\ref {Eisenstein, D. J. and Loeb, A. 1996,  \apj{459}{432}}
 
\ref{Giovanelli, R. \etal 
1997, \aj{}{in press}}

\ref{de Grijs, R. and van der Kruit, P.C. 1996, \aas{117}{19}}

\ref{Elbaz, D., Arnaud, M and Vangioni-Flam, E. 1995, \aa{303}{345}}

\ref{Fisher, D., Franx, M. and  Illingworth, G. 1995, \apj{448}{119}} 

\ref{Kauffmann, G., Guiderdoni, B., and White, S. D. M.  1994,\mn{267}{981}}

 \ref{Kauffmann, G., White, S. D. M. and Guiderdoni, B. 1993,\mn{264}{201}}

\ref{Kennicutt, R. C.  1989, \apj{344}{685}}

\ref{Kennicutt, R. C., Tamblyn, P. and  Congdon, C. W. 1994, \apj{435}{22}} 

\ref{Kent, S.M. 1987, \aa{93}{816}}
    
\ref{ Kent, S.M. 1988, \aa{96}{514}}

\ref{Lacey, C. and  Silk, J. 1991, \apj{381}{14}}

\ref{Lacey, C., Guiderdoni, B., Rocca-Volmerange, B. and  Silk, J. 1993,
\apj{402}{15}}

\ref{Le Brun, V.,  Bergeron, J. and Boisse, P. 1996, \aa{306}{691}}
     
\ref{McKee, C. F. 1989, \apj{345}{782}}

\ref{Navarro, J. and Steinmetz, M. 1997, \apj {}{in press}}

\ref{Noh, H. and Scalo, J. 1990, \apj{352}{605}}

\ref{Prantzos, N.  and Aubert, O. 1995, \aa{302}{69}}

\ref{Rees, M. J. and Ostriker, J. P. 1977, \mn{179}{451}} 

\ref{Renzini, A., Ciotti, L., D'Ercole, A. and Pellegrini, S. 1993, \apj{419}{52}}

\ref{Shelton, R. L.  and Cox, D. P.  1994, \apj{434}{599}}

\ref{Shull, J.M., Stocke, J.T. and Penton, S. 1996, \aj{1112}{72}}

\ref{Silk, J.  1977, \apj{211}{638}}

\ref{Silk, J. and Wyse, R. F. G.  1997, in preparation.}

\ref{Steidel, C.C., Pettini, M., Dickinson, M. and Persson, S.E.
     1994, \aj{108}{2046}}

\ref{Stark, A. A.  and Brand, J.  1989, \apj{339}{763}}

\ref{ Steinmetz, M. and Muller, E. 1995, \mn{89}{45}}

\ref{Strauss, M. A. and Willick, J. A. 1995, \phyr{261}{272}}

\ref{Thoul, A. A. and  Weinberg, D. H. 1995, \apj{442}{480}}

\ref{Wang, B.  and Silk, J.  1993, \apj{427}{759}}

\ref{ Willick, J. A. 1997, in  {\it Formation of Structure in the Universe}, (eds. A. Dekel and J. P. Ostriker) (Cambridge University Press, Cambridge), in press}

\ref{ Wyse, R. F. G. and Silk, J. 1989, \apj{339}{700}} 

\ref{ Zepf, S.  and Silk, J. 1996, \apj{446}{114}} 

\ref{Zwaan, M. A. \etal 
1995, \mn{273}{L35}}

\bye